\documentclass[showkeys]{revtex4}

\usepackage{epsfig}

\begin{document}

\title{Unified Approach to the Large-Signal and High-Frequency Theory
of $p\!-\!n$-Junctions}

\author{Anatoly A. Barybin$^{*}$, and Edval J. P. Santos$^{**}$}
\affiliation{$^{*}$Electronics Department,
Saint-Petersburg State Electrotechnical University, 
197376, Saint-Petersburg, Russia\\
$^{**}$Laboratory for Devices and Nanostructures,
Departamento de Eletr\^{o}nica e Sistemas,\\
Universidade Federal de Pernambuco,\\
C.P. 7800, 50670-000, Recife-PE, Brasil\\
E-mail: edval@ee.ufpe.br}

\begin{abstract}
Spectral approach to the charge carrier transport in $p\!-\!n$-junctions
has allowed us to revise the theoretical results relating to large
signal operation of the junction to make them valid for both the low and
high frequencies ranges.
The spectral composition of the external circuit current
includes both the DC and AC components. The former
produces the static current--voltage characteristic with the modified
Bessel function of zeroth order, $I_0(\beta V_\sim)$, while the latter
yields the dynamic admittance with the modified Bessel
function of first order, $I_1(\beta V_\sim)$, either of them depend on
the signal amplitude.
Our experimental results are consistent with the theoretical ones if a
fitting parameter allowing for recombination processes in the depletion
layer is taken into account.

\end{abstract}


\maketitle

\section{Introduction}
\label{sec:1}

The conventional theory of $p\!-\!n$-junctions is usually studied
in two limiting cases: small-signal/high-frequency and
large-signal/low-frequency~\cite{1,2,3,4,5,6}. We have found
out that it is possible to unify these two cases by
applying the spectral analysis to calculating
the external circuit current for a $p\!-\!n$-diode. General scheme
of finding the spectral solution to the diffusion equations for
the $p\!-\!n$-junction under arbitrary signal action is developed in
Sec.~\ref{sec:2} and then applied in Sec.~\ref{sec:3} to calculate
the external circuit current. For this work, it is assumed that 
the relaxation times, and the diffusion constants do not change 
with injection level, or constant effective values are used. 
The nonlinearity considered in this work is related to the
doping profile, as it yields a semiconducting $p\!-\!n$-juntion.
Sections~\ref{sec:4}
and~\ref{sec:5} deal with the static current--voltage and dynamic
characteristics of the diode. First measurements to validate 
the experimental setup is set forth in Sec.~\ref{sec:6}.
Appendix examines the
conventional approach (based on applying the customary current--voltage
characteristic expression to time-varying conditions) to be compared
with our results obtained below in terms of the spectral approach.

\section{Spectral Solution of Diffusion Equations}
\label{sec:2}

Theoretical basis of the charge carrier transport
in $p\!-\!n$-junctions is the diffusion--drift equations for minority
carriers in the $n$- and $p$-regions. These equations are derived
from the continuity equations together~with the current density
expressions for holes and electrons. Such diffusion--drift equations
have the following one-dimensional form~\cite{4,5}:
\vskip2pt
(a) for holes (with equilibrium and nonequilibrium densities $p_n$
and $p(z,t)\,$) injected into the $n$-region
\begin{equation}
{\partial p\over\partial t} + {\partial(p\mu_p E)\over\partial z} -
{\partial^2(D_p p)\over\partial z^2} + \frac{p -p_n}{\tau_p} = 0 ,
\label{eq:1}
\end{equation}

(b) for electrons (with equilibrium and nonequilibrium densities $n_p$
and $n(z,t)\,$) injected into the $p$-region
\begin{equation}
{\partial n\over\partial t} - {\partial(n\mu_n E)\over\partial z} -
{\partial^2(D_n n)\over\partial z^2} + \frac{n -n_p}{\tau_n} = 0 .
\label{eq:2}
\end{equation}

We shall apply Eqs.~(\ref{eq:1}) and (\ref{eq:2}) to study the simplest
Shockley's model of the ideal diode~\cite{1} but operating at the nonlinear
and multifrequency regime under action of the harmonic signal of arbitrary
amplitude. In the case of low-level injection in $p\!-\!n$-diodes, one
conventionally assumes $E=0$ for the neutral parts of the $p$- and
$n$-regions~\cite{4,5}. Then Eqs.~(\ref{eq:1}) and (\ref{eq:2}) turn
into the diffusion equations for the excess concentrations (over their
equilibrium values $p_n$ and $n_p$) for holes, ${\Delta p(z,t)=p(z,t)-p_n}$,
and for electrons, $\Delta n(z,t)=n(z,t)-n_p$, of the following form:
\begin{equation}
\biggl( 1+ \tau_p{\partial\over\partial t} \,\biggr) \Delta p -
L_p^2\, {\partial^2\Delta p\over\partial z^2} = 0,  \;\;
\label{eq:3}
\end{equation}
\begin{equation}
\biggl( 1+ \tau_n{\partial\over\partial t} \,\biggr) \Delta n -
L_n^2\, {\partial^2\Delta n\over\partial z^2} = 0,
\label{eq:4}
\end{equation}
where $L_p=\sqrt{D_p\tau_p}$ and $L_n=\sqrt{D_n\tau_n}$ are the
diffusion lengths for holes and electrons, respectively.

For the resonant circuit connection, the total voltage applied to the
$p\!-\!n$-diode consists of the DC bias voltage $V_0$ and the harmonic
signal voltage $V_\sim \cos\omega t$:
\begin{equation}
v(t)= V_0 + V_\sim \cos\omega t.
\label{eq:5}
\end{equation}

Nonlinearity of electronic processes in the $p\!-\!n$-junction produces
frequency harmonics $k\omega$ so that the required solutions of
Eqs.~(\ref{eq:3}) and (\ref{eq:4}) can be represented in the complex
form of Fourier series:
\begin{equation}
\Delta p(z,t)=\!\!
\sum_{k\,=-\infty}^{\infty} \!\Delta p_k(z)\,e^{ik\omega t},
\label{eq:6}
\end{equation}
\begin{equation}
\Delta n(z,t)=\!\!
\sum_{k\,=-\infty}^{\infty} \!\Delta n_k(z)\,e^{ik\omega t}.
\label{eq:7}
\end{equation}

Real values of $\Delta p(z,t)$ and $\Delta n(z,t)$ are provided with
the following relations for the complex amplitudes:\,
$\Delta p_k=\Delta p_{-k}^*$ \,and\, $\Delta n_k=\Delta n_{-k}^*$.

Substitution of the required solutions (\ref{eq:6}) and (\ref{eq:7})
into Eqs.~(\ref{eq:3}) and (\ref{eq:4}) with taking into account
the orthogonality of harmonics reduces to the following equations for
the desired complex amplitudes of harmonics:
\begin{equation}
{d^2\Delta p_k\over dz^2} - \frac{\Delta p_k}{L_{pk}^2} = 0
\;\quad\mbox{and}\quad\;
{d^2\Delta n_k\over dz^2} - \frac{\Delta n_k}{L_{nk}^2} = 0
\label{eq:8}
\end{equation}
with
\[
L_{pk}=\frac{L_p}{\sqrt{1+ik\omega \tau_p}}
\;\quad\mbox{and}\quad\;
L_{nk}=\frac{L_n}{\sqrt{1+ik\omega \tau_n}}\,.
\]

General solutions of equations (\ref{eq:8}) for the $p\!-\!n$-diode
with the junction region $-W_p<z<W_n$ have the following form: \\
\indent
(a) for holes injected into the $n$-region ($z\geq W_n$)
\begin{eqnarray}
\Delta p_k(z) =
C_{pk}^+ \exp \biggl( \!-\Lambda_{pk}\frac{z-W_n}{L_p} \biggr) +
C_{pk}^- \exp \biggl( \Lambda_{pk}\frac{z-W_n}{L_p} \biggr) ,
\label{eq:9}
\end{eqnarray}

where $\Lambda_{pk} = a_{pk} + i b_{pk}$ \,and
\begin{equation}
a_{pk}= \frac{1}{\sqrt{2}}\,\sqrt{1+\sqrt{1+(k\omega\tau_p)^2} }\,, \qquad
b_{pk}=\frac{k\omega\tau_p}{2a_{pk}}\,;
\label{eq:10}
\end{equation}

(b) for electrons injected into the $p$-region ($z\leq -W_p$)
\begin{eqnarray}
\Delta n_k(z) =
C_{nk}^+ \exp \biggl( \!-\Lambda_{nk}\frac{z+W_p}{L_n} \biggr) +
C_{nk}^- \exp \biggl( \Lambda_{nk}\frac{z+W_p}{L_n} \biggr) ,
\label{eq:11}
\end{eqnarray}

where $\Lambda_{nk}=a_{nk}+ ib_{nk}$\, and
\begin{equation}
a_{nk}= \frac{1}{\sqrt{2}}\,\sqrt{1+\sqrt{1+(k\omega\tau_n)^2} }\,,  \qquad
b_{nk}=\frac{k\omega\tau_n}{2a_{nk}}\,.
\label{eq:12}
\end{equation}

According to Eqs.~(\ref{eq:10}) and (\ref{eq:12}), always
Re$\Lambda_{pk}= a_{pk}>0$ \,and\, Re$\Lambda_{nk}=a_{nk}>0$. So for the
$p\!-\!n$-diodes with thick base (when $d_n\gg L_p$ and $d_p\gg L_n$,
where $d_n$ and $d_p$ are thicknesses of the neutral parts) we can take
$C_{pk}^- =0$ and $C_{nk}^+ =0$. This eliminates the necessity for
boundary conditions on ohmic contacts. Taking into account (\ref{eq:9})
and (\ref{eq:11}) with $C_{pk}^-= C_{nk}^+=0$, the general solutions
(\ref{eq:6}) and (\ref{eq:7}) of Eqs.~(\ref{eq:3}) and (\ref{eq:4})
assume the form \\[-2mm]
\begin{equation}
\Delta p(z,t)=\! \sum_{k\,=-\infty}^{\infty} \!C_{pk}^+
\exp \biggl( \!-\Lambda_{pk}\frac{z-W_n}{L_p} \biggr) \,e^{ik\omega t} ,
\label{eq:13}
\end{equation}
\begin{equation}
\Delta n(z,t)=\! \sum_{k\,=-\infty}^{\infty} \!C_{nk}^-
\exp \biggl( \Lambda_{nk}\frac{z+W_p}{L_n} \biggr) \,e^{ik\omega t} .
\label{eq:14}
\end{equation}

These expressions provide for $\Delta p(z,t)\to 0$ as $z\to\infty$ and
$\Delta n(z,t)\to 0$ as $z\to -\infty$ because of Re$\Lambda_{pk}>0$
and Re$\Lambda_{nk}>0$.

The constants $C_{pk}^+$ and $C_{nk}^-$ in Eqs.~(\ref{eq:13})
and (\ref{eq:14}) can be found from the conventional injection boundary
conditions~\cite{4,5}:
\begin{equation}
\Delta p(W_n,t)= p_n f(t) \quad\qquad \mbox{for} \quad z=W_n,
\label{eq:15}
\end{equation}
\begin{equation}
\Delta n(-W_p,t)= n_p f(t) \qquad \,\mbox{for} \quad z=-W_p,
\label{eq:16}
\end{equation}
where for the applied voltage $v(t)$ of the form (5) we have
introduced the function
\begin{equation}
f(t) = \exp \biggl( \frac{qv(t)}{\kappa T} \biggr) - 1.
\label{eq:17}
\end{equation}
where $\kappa= 1.38 \times 10^{-23}\,\,J/K$ is the Bolztmann constant,
and $T$ is the temperature in kelvin.

Substitution of (\ref{eq:13}) and (\ref{eq:14}) into the
boundary conditions~(\ref{eq:15}) and (\ref{eq:16}) yields
\begin{equation}
\sum_{k\,=-\infty}^{\infty} \!C_{pk}^+ \,e^{ik\omega t} = p_n f(t),
\label{eq:18}
\end{equation}
\begin{equation}
\sum_{k\,=-\infty}^{\infty} \!C_{nk}^- \,e^{ik\omega t} = n_p f(t).
\label{eq:19}
\end{equation}

By using the orthogonality property of harmonics in the expansions
(\ref{eq:18}) and (\ref{eq:19}), it is easy to obtain the desired
constants:
\begin{equation}
C_{pk}^+ = p_n F_k  \qquad\mbox{and}\qquad
C_{nk}^- = n_p F_k,
\label{eq:20}
\end{equation}
where $F_k$ is the Fourier amplitude of the $k$th harmonic for the
function $f(t)$ given by Eq.~(\ref{eq:17}), that is
\begin{equation}
F_k = \frac{1}{2\pi} \int\limits_{-\pi}^{\;\pi}
f(t) e^{-ik\omega t} \,d\omega t .
\label{eq:21}
\end{equation}
Substitution of the function (\ref{eq:17}) into formula (\ref{eq:21}) gives
\begin{equation}
F_0 = I_0(\beta V_{\sim})\exp (\beta V_0) - 1
\quad\quad\;\;\mbox{for}\quad       k=0,
\label{eq:22}
\end{equation}
\begin{equation}
F_k= F_{-k} = I_k(\beta V_{\sim})\exp (\beta V_0)
\quad\;\mbox{for}\quad              k\neq 0.
\label{eq:23}
\end{equation}

Here we have introduced the modified Bessel functions of the first kind
of order $k$ ($k=0,\pm 1,\pm 2, \ldots$) which have the following
integral representation (see, for example, formula~8.431.5
in~\cite{7}):
\begin{equation}
I_k(\beta V_{\sim})=\frac{1}{\pi}
\int\limits_0^{\;\pi} e^{ \beta V_{\sim}\cos\omega t}
\cos k\omega t \:d\omega t.
\label{eq:24}
\end{equation}
These functions depend on $\beta V_{\sim}$, where $V_{\sim}$ is an
amplitude of the signal applied to the $p\!-\!n$-junction and
$\beta = q/\kappa T$.

Consequently, with making allowance for (\ref{eq:20}) the general
solutions (\ref{eq:13}) and (\ref{eq:14}) of the diffusion
equations~(\ref{eq:3}) and (\ref{eq:4}) take the final form of
spectral expansions:
\begin{equation}
\Delta p(z,t)=\, p_n\! \sum_{k\,=-\infty}^{\infty}\!
F_k \exp \biggl( \!-\Lambda_{pk}\frac{z-W_n}{L_p} \biggr)\, e^{ik\omega t} ,
\label{eq:25}
\end{equation}
\begin{equation}
\Delta n(z,t)=\, n_p\! \sum _{k\,=-\infty}^{\infty}\!
F_k \exp \biggl( \Lambda_{nk}\frac{z+W_p}{L_n} \biggr)\, e^{ik\omega t} .
\label{eq:26}
\end{equation}

These expressions allow us to obtain the spectral composition of
the current flowing through an external circuit connected to the
$p\!-\!n$-diode.

\section{External Circuit Current for $p\!-\!n$-Diodes}
\label{sec:3}

The initial equation to derive an expression for the diode current
is the total current conservation law of the general form
$\nabla\cdot(\,{\bf j}_p+\,{\bf j}_n +\,
\epsilon {\partial{\bf E}/\partial t})=0$\, following from Maxwell's
equations. Its one-dimensional form (along the $z$-direction) is
\begin{equation}
{\partial\over\partial z} \biggl( j_{pz} + j_{nz} +
\epsilon\!\:{\partial E_z\over\partial t} \biggr) = 0.
\label{eq:27}
\end{equation}

The quantity in parentheses of Eq.~(\ref{eq:27}), being independent of~$z$,
defines the external circuit current (for the cross-section area $S$ of the diode)
as a function of time:
\begin{eqnarray}
J(t)
&\equiv&
j_{pz}(z,t)S + j_{nz}(z,t)S +
\epsilon\!\:{\partial E_z(z,t)\over\partial t}\!\:S
\nonumber\\[2mm]
&=&
q(\mu_p p + \mu_n n) E_z S -
q \biggl(\! D_p{\partial p\over\partial z} -
         D_n {\partial n\over\partial z} \biggr)S +
\epsilon\!\:{\partial E_z\over\partial t}\!\:S \,.  \qquad
\label{eq:28}
\end{eqnarray}

We have used the usual expressions for the hole and electron current
densities~\cite{5} involving the drift and diffusion components which
underlie the initial diffusion--drift equations~(\ref{eq:1}) and
(\ref{eq:2}). All the terms on the right of Eq.~(\ref{eq:28})
depend on both $z$ and~$t$, but taken together in any cross section
they yield the external circuit current as a function of only time.

Restricting our consideration to the $p\!-\!n$-diodes with low injection
($p\ll n_n$ but $p> p_n$ so that the excess hole density is 
$\Delta p=p-p_n>0$, for $n$-type, similar conditions hold for $p$-type),
as was noted above, we can assume $E_z=0$ in neutral parts of the $p$- and
$n$-regions~\cite{5}. Then the external circuit current (\ref{eq:28}) is
determined only by the diffusion currents taken at any cross section,
for example, at $z=W_n$:
\begin{eqnarray}
J(t) &=&
- q D_p {\partial p(z,t)\over\partial z} \,\bigg|_{z\,=\,W_n}\!\!S \;+\;
q D_n {\partial n(z,t)\over\partial z} \,\bigg|_{z\,=\,W_n}\!\!S
\nonumber\\[2mm]
&\equiv&  \,j_{pz}(W_n,t)S\,+\,j_{nz}(W_n,t)S.
\label{eq:29}
\end{eqnarray}

Neglecting recombination processes inside the $p\!-\!n$-junction (which
is true if $W_n\ll L_p$ \,and\, $W_p\ll L_n$) we~can write~\cite{5}
\begin{equation}
j_{nz}(W_n,t) = j_{nz}(-W_p,t).
\label{eq:30}
\end{equation}

Substitution of (\ref{eq:30}) into Eq.~(\ref{eq:29}) gives the external
circuit current:
\begin{eqnarray}
J(t) &=&
j_{pz}(W_n,t)S \,+\, j_{nz}(-W_p,t)S
\nonumber\\[1.5mm]
&=&\!
-q D_p {\partial\Delta p(z,t)\over\partial z}\,
\bigg|_{z\,=\,W_n}\!\!S \;+\; q D_n {\partial\Delta n(z,t)\over\partial z}\,
\bigg|_{z\,=\,-W_p}\!\!S, \qquad
\label{eq:31}
\end{eqnarray}
where $\Delta p = p -p_n$ and $\Delta n = n -n_p$ are the excess
concentrations of injected carriers given by Eqs.~(\ref{eq:25}) and
(\ref{eq:26}). Inserting these formulas into expression (\ref{eq:31}),
we finally obtain the spectral representation for the external circuit
current:
\begin{equation}
J(t) = J_{s,p}\! \sum_{k\,= -\infty}^{\infty}\!
F_k\!\:\Lambda_{pk}\, e^{ik\omega t} \,+\,
J_{s,n}\! \sum_{k\,= -\infty}^{\infty}\!
F_k\!\:\Lambda_{nk}\, e^{ik\omega t}.
\label{eq:32}
\end{equation}

Here, the hole and electron contributions into the saturation current
of the thick $p\!-\!n$-diode are defined, as it is generally
accepted~\cite{4,5}, in the following form:
\begin{equation}
J_{s,p} = \frac{qp_n D_p}{L_p}\,S  \,\;\quad\mbox{and}\quad\;
J_{s,n} = \frac{qn_p D_n}{L_n}\,S.
\label{eq:33}
\end{equation}

Expression (\ref{eq:32}) contains all the spectral components of the external
circuit current, among which the terms with numbers $k= 0$
(the DC current) and $k=\pm 1$ (the AC current) are of most interest for our
subsequent consideration.

\section{Static Current--Voltage Characteristic of $p\!-\!n$-Diodes}
\label{sec:4}

The term in series (\ref{eq:32}) numbered by $k=0$ corresponds to the DC
current $J_0$, which by using expression~(\ref{eq:22}) for $F_0$ and
$\Lambda_{p0}=\Lambda_{n0}=1$ can be written in the form of the
{\it static current--voltage characteristic}:
\begin{equation}
J_0(V_0,V_{\sim})=
J_s \Bigl[\!\: I_0(\beta V_{\sim})\,e^{\beta V_0} -\!1 \!\:\Bigr].
\label{eq:34}
\end{equation}
Here, $I_0(\beta V_{\sim})$ is the modified
Bessel function of zeroth order depending on the signal amplitude
$V_{\sim}$. The saturation current $J_s$ has the customary
form~\cite{4,5}
\begin{equation}
J_s = J_{s,p} + J_{s,n} =
\frac{qp_n D_p}{L_p}\,S + \frac{qn_p D_n}{L_n}\,S.
\label{eq:35}
\end{equation}

Expression (\ref{eq:34}) differs in appearing the Bessel function
$I_0(\beta V_{\sim})$ from the customary current--voltage characteristic
given in such well-known books as, for example,~\cite{4,5}. Coincidence
between them takes place only when $I_0(\beta V_{\sim})\simeq 1$,
which is true for small signals with $V_{\sim}\!\ll\kappa T/q$.
However, an expression similar to our equation~(\ref{eq:34}) was found
in a less-known textbook~\cite{6}\,\footnote{The authors are grateful to
a referee for indicating this reference.} written in German, whose
derivation is adduced for comparison in Appendix
(see formula~(\ref{eq:A.4})).

The modified Bessel function ${I_0(\beta V_{\sim})\geq 1}$ as a function
of the signal amplitude~$V_{\sim}$ mathematically reflects the so-called effect
of {\it signal rectification\/}. This effect provides a~contribution into
the DC current $J_0$ from the signal and results in up-shifts of the
static current--voltage characteristic $J_0(V_0)$ with increasing
$V_{\sim}$, as shown in Fig.~1.

\begin{figure}
\epsfysize=3in
\centerline{\epsffile{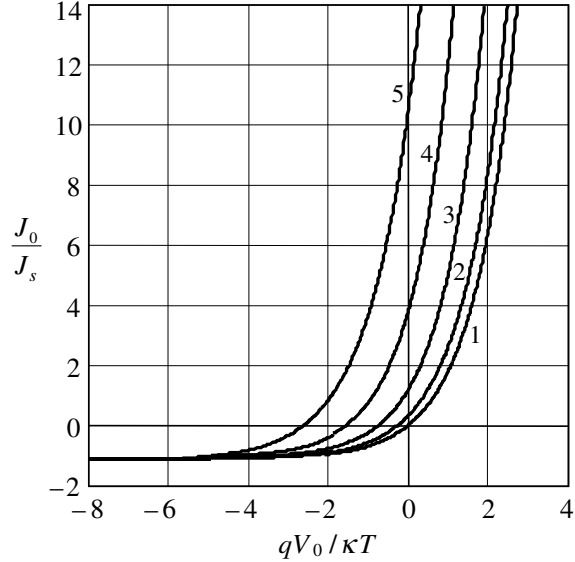}}
\caption{Static current--voltage characteristics for different
values of the signal voltage amplitudes: ${V_{\sim}=0}$ (curve 1),
$\kappa T/q$ (curve 2), $2\kappa T/q$ (curve 3), $3\kappa T/q$ (curve 4),
$4\kappa T/q$ (curve 5).}
\label{Fig1}
\end{figure}

\section{Dynamic Characteristics of $p\!-\!n$-Diodes}
\label{sec:5}

The first harmonic current contribution to the general expression (\ref{eq:32})
for the external current corresponds to terms numbered by $k= \pm 1$ and is
equal to
\begin{equation}
J_1(t) = J_1 e^{i\omega t} +\, c.c.
\;\quad\mbox{with}\quad\;
J_1 = F_1 (J_{s,p}\Lambda_{p1} + J_{s,n}\Lambda_{n1}),
\label{eq:36}
\end{equation}
where the quantities $\Lambda_{p1}$, $ \Lambda_{n1}$, $F_1$ are
respectively given by Eqs. (\ref{eq:10}), (\ref{eq:12}),
(\ref{eq:23}) for~$k=1$.

Expressions (\ref{eq:5}) with ${V_\sim=2V_1}$ and (\ref{eq:36}) allow one to
introduce for the $p\!-\!n$-diode its {\it dynamic admittance\/} defined as
\begin{equation}
Y = \frac{J_1}{V_1} \equiv G_d + i\omega C_d .
\label{eq:37}
\end{equation}

Here, the {\it dynamic conductance\/} $G_d$ and
the {\it dynamic\/} ({\it diffusion\/}) {\it capacitance\/} $C_d$ are
introduced in a customary way~\cite{5}. After substituting (\ref{eq:36})
into (\ref{eq:37}) and some transformations we obtain these quantities
as functions of frequency:
\begin{equation}
G_d(\omega) = g G_0 \biggl(
a_{p1}(\omega){J_{s,p}\over J_s} \!\:+\!\:
a_{n1}(\omega){J_{s,n}\over J_s} \biggr) ,
\label{eq:38}
\end{equation}
\begin{equation}
C_d(\omega) = {g G_0\over\omega} \biggl(
b_{p1}(\omega){J_{s,p}\over J_s} \!\:+\!\:
b_{n1}(\omega){J_{s,n}\over J_s} \biggr) ,  \,
\label{eq:39}
\end{equation}
where the quantities $a_{p1},b_{p1}$ and $a_{n1},b_{n1}$ follow from
Eqs. (\ref{eq:10}) and (\ref{eq:12}) with $k=1$.

Formulas (\ref{eq:38}) and (\ref{eq:39}) have been derived by inserting
the differential conductance introduced from the static current--voltage
characteristic $J_0(V_0, V_\sim)$ defined as
\begin{equation}
G_0 = {\partial J_0(V_0, V_\sim)\over\partial V_0} \,.
\label{eq:40}
\end{equation}

The static current--voltage characteristic, Eq.~(\ref{eq:34}),
gives the differential~conduc\-tance (\ref{eq:40}) which depends on
both the bias voltage $V_0$ and the signal amplitude~$V_\sim$:
\begin{equation}
G_0(V_0,V_{\sim}) = \frac{J_0 + J_s}{\kappa T/q} =\,
{J_s\exp(\beta V_0)\over \kappa T/q} \,I_0(\beta V_\sim) \,.
\label{eq:41}
\end{equation}

Formulas (\ref{eq:38}) and (\ref{eq:39}) contain a combination of the
modified Bessel functions $I_0(\beta V_{\sim})$ and $I_1(\beta V_{\sim})$
in the form of a factor:
\begin{equation}
g(V_\sim) = {2\over\beta V_{\sim}}\,
{I_1(\beta V_{\sim})\over I_0(\beta V_\sim)}\equiv
\frac{g_1(V_\sim)}{I_0(\beta V_\sim)} \leq 1
\label{eq:42}
\end{equation}
with
\begin{equation}
g_1(V_\sim) =
{I_1(\beta V_{\sim})\over\beta V_\sim/2}\,.
\label{eq:43}
\end{equation}

These quantities depend on the signal amplitude $V_\sim$, so
that $g(V_\sim)\simeq g_1(V_\sim)\simeq 1$ for small signals when
${V_{\sim}\ll\kappa T/q}$ \,and\, $g(V_\sim)\simeq 2/(\beta V_\sim)\!\to\!0$
\,as\, $V_\sim\!\to\!\infty$, whereas here $g_1(V_\sim)$ is approximated
by an exponentially growing function.

After inserting (\ref{eq:41}) and (\ref{eq:42}) into
Eqs.~(\ref{eq:38}) and (\ref{eq:39}), they take the following form:
\begin{equation}
G_d(\omega) = g_1 G_{d0} \,\biggl(
a_{p1}(\omega) {J_{s,p}\over J_s} \!\:+\!\:
a_{n1}(\omega) {J_{s,n}\over J_s} \biggr),
\label{eq:44}   \\[2mm]
\end{equation}
\begin{equation}
C_d(\omega) = g_1 C_{d0} \,\biggl(
a_{p1}^{-1}(\omega) {Q_p\over Q} \!\:+\!\:
a_{n1}^{-1}(\omega) {Q_n\over Q} \biggr),
\label{eq:45}
\end{equation}
where\, $Q=Q_p+Q_n$ \,with\, $Q_p\equiv J_{s,p}\!\:\tau_p = qp_n L_p S$ \,and\,
$Q_n\equiv J_{s,n}\!\:\tau_n = qn_p L_n S$.
The quantities $G_{d0}$ and $C_{d0}$ appearing in Eqs.~(\ref{eq:44}) and
(\ref{eq:45}) as functions of $V_0$ are equal to
\begin{eqnarray}
G_{d0}(V_0) =
{J_s\exp(\beta V_0)\over\kappa T/q} =
{q S\over \kappa T} \biggl( \frac{qp_n D_p}{L_p} +
\frac{qn_p D_n}{L_n} \biggr)\!\:e^{qV_0/ \kappa T} ,
\label{eq:46}
\end{eqnarray}
\begin{eqnarray}
C_{d0}(V_0) =
{Q\exp(\beta V_0)\over2\kappa T/q} =
{q S\over\kappa T} \biggl( {qp_n L_p\over2} +
{qn_p L_n\over2} \biggr)\!\:e^{qV_0/\kappa T}.
\label{eq:47}
\end{eqnarray}

Expressions (\ref{eq:46}) and (\ref{eq:47}) correspond to the diffusion
conductance and capacitance at low frequencies (when $\omega\tau_{p,n}\ll1$
and $a_{p1}(\omega)=a_{n1}(\omega)\simeq1$) for small signals
(when $V_{\sim}\ll\kappa T/q$ and $g_1(V_{\sim})\simeq1$).
Our expressions for $G_{d0}$ and $C_{d0}$ fully coincide with those
given in a book~\cite{5} by Sze (see there formulas (65) and (66) of Chapter~2).

As seen from (\ref{eq:44}) and (\ref{eq:45}), the frequency dependence
of the dynamic conductance $G_d(\omega)$ and the diffusion capacitance
$C_d(\omega)$ is determined by the functions $a_{p1}(\omega)$ and
$a_{n1}(\omega)$ given by formulas (\ref{eq:10}) and (\ref{eq:12}) with
${k=1}$, whereas their dependence on the signal amplitude $V_{\sim}$ is
described by the function $g_1(V_\sim)$ introduced in the form of
expression~(\ref{eq:43}).

A certain simplification of expressions~(\ref{eq:44}) and (\ref{eq:45})
occurs in the case of the {\it one-sided\/} $p^+\!-n$-junction with highly
doped emitter when ${p_n\gg n_p}$,\, ${J_{s,p}\gg J_{s,n}}$,\, ${Q_p\gg Q_n}$
so that $J_s\simeq J_{s,p}$ and $Q\simeq Q_p$. Then
\begin{equation}
{G_d(\omega,V_{\sim})\over G_{d0}(V_0)} =
{ \sqrt{1+\sqrt{1+ \omega^2\tau_p^2}}\over\sqrt{2}}\;
{\,I_1(\beta V_{\sim})\over \beta V_{\sim}/2} \,,
\label{eq:48}
\end{equation}
\begin{equation}
{C_d(\omega,V_{\sim})\over C_{d0}(V_0)} =
{ \sqrt{2}\over\sqrt{1+\sqrt{1+ \omega^2\tau_p^2}}}\;
{\,I_1(\beta V_{\sim})\over \beta V_{\sim}/2} \,,
\label{eq:49}
\end{equation}
where from Eqs.~(\ref{eq:46}) and (\ref{eq:47}) it follows that
\begin{equation}
G_{d0}(V_0) = \frac{2}{\tau_p}\,C_{d0}(V_0) =
{J_s\exp(\beta V_0)\over\kappa T/q}\,.
\label{eq:50}
\end{equation}

Frequency dependencies given by expressions (\ref{eq:48}) and
(\ref{eq:49}) are plotted in Fig.~2. The plots differ from the similar
curves shown by Sze~\cite{5} (see there Fig.~23 of Chapter~2) in that
they take into account the signal amplitude influence owing to the
factor ${g_1(V_\sim)=I_1(\beta V_\sim)/(\beta V_\sim /2)}$. For small
signals (when ${\beta V_\sim\ll 1}$ and $g_1(V_\sim)\simeq 1$) our
expressions (\ref{eq:48}) and (\ref{eq:49}) assume the form obtained by
Sze~\cite{5} and are depicted by curves~1 in Fig.~2. The low-frequency
values (when $\omega\tau_p\ll 1$) of the quantities (\ref{eq:48}) and
(\ref{eq:49}) for arbitrary signals are equal to
\[
{G_d(0,V_{\sim})\over G_{d0}(V_0)} =
{C_d(0,V_{\sim})\over C_{d0}(V_0)} =
{I_1(\beta V_{\sim})\over \beta V_{\sim}/2}\equiv g_1(V_\sim) .
\]

\begin{figure}
\epsfysize=3.3in
\centerline{\epsffile{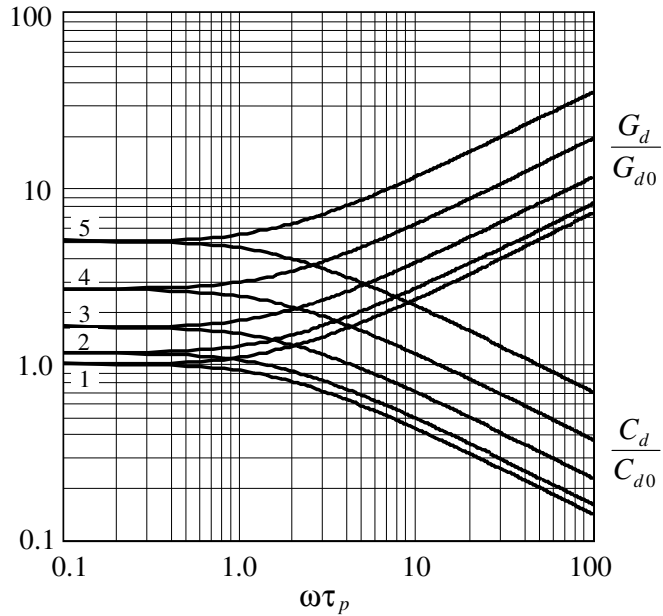}}
\label{Fig2}
\caption{Frequency dependencies of the normalized dynamic conductance
$G_d(\omega)/G_{d0}$ and diffusion capacitance $C_d(\omega)/C_{d0}$
for different values of the signal voltage amplitudes: ${V_{\sim}=0}$
(curve 1), $\kappa T/q$ (curve 2), $2\kappa T/q$ (curve 3), $3\kappa T/q$
(curve 4), $4\kappa T/q$ (curve 5).}
\end{figure}

From here it follows that the low-frequency value (denoted by zero in
parentheses that means $\omega=0$) of the dynamic conductance is described
by the following expression:
\begin{equation}
G_d(0,V_{\sim})= G_{d0}(V_0)\,g_1(V_\sim).
\label{eq:51}
\end{equation}

Eq.\ (\ref{eq:51}) coincides with formula (\ref{eq:A.5}) obtained
from the conventional approach based on using the customary
current--voltage characteristic for time-varying conditions.

The conventional approach is not able to provide the frequency dependence
not only for $G_d(\omega)$ but also for $C_d(\omega)$ such as given by
Eq.~(\ref{eq:48}) and (\ref{eq:49}) which are valid for signals of
arbitrary amplitude. The similar frequency dependency known from the
published literature was obtained only for small signals when
$V_{\sim}\!\ll\!\kappa T/q$. Such a result can be found,
for example, in Sec.~2.4.4 of a book by Sze~[5], where the curves
$G_d(\omega)/G_{d0}$ and $C_d(\omega)/C_{d0}$ given in Fig.~23
correspond to our curves~1 in Fig.~2 calculated for $V_{\sim}=0$.
Other our curves denoted by numbers~2, 3, 4, 5 in Fig.~2 are novel
since they demonstrate a dependence of the dynamic conductance
and capacitance on the AC signal amplitude~$V_{\sim}$, which was
previously missed. It is this dependence that is of interest to check
experimentally.

\section{Experimental Validation}
\label{sec:6}

The above theoretical results have been derived from the ideal
Shockley's model which have a few simplifying assumptions. Among them
the most essential ones are: (i)~the large thickness of the neutral
regions as compared with the minority-carrier diffusion lengths, i.~e.,
${d_n\gg L_p}$ and ${d_p\gg L_n}$; (ii)~the neglect of recombination
effects in the depletion layers owing to inequalities $W_n\ll L_p$ and
$W_p\ll L_n$. Yet, such assumptions are rarely verified in practical
conditions and should be taken into account, as applied to real
devices used in our experiments.

The practically used $p^+\!\!-n$-diode with the one-sided injection
has the thin $n$-base of thickness $d_n$ compared to~$L_p$. In this
case, it is necessary to apply the boundary condition taking into
account surface recombination on a metallic contact of the base
located at $z=W_n+d_n$ (see Eq.~(108) of Chapter~1 in Ref.~\cite{5}):
\[
\frac{\partial\Delta p(z,t)}{\partial z} \biggl|_{z\,=\,W_n+\,d_n} \!=\,
-\frac{s_p}{D_p}\, \Delta p(W_n\!+d_n,t),
\]
where $s_p$ is the speed of surface recombination of holes (for ohmic
contacts $s_p\!\to\!\infty$). The additional boundary condition changes
only the saturation current by the factor~$A_p$:
\begin{equation}
J_s = \frac{q p_n D_p}{L_p}\,S A_p
\;\quad\mbox{with}\quad\;
A_p = \frac{(s_p L_p/D_p) +
\tanh((W_n + d_n)/L_p)}{1+(s_p L_p/D_p)\tanh((W_n + d_n)/L_p)}\,.
\label{eq:52}
\end{equation}

As follows from Eq.~(\ref{eq:50}), the measurement of the small-signal
low-frequency conductance $G_{d0}=\beta J_s\exp(\beta V_0)$ will give
us the saturation current (\ref{eq:52}) with allowance for the
{\it surface recombination\/} on metallic contact of the base.

To take into account the {\it bulk recombination\/} processes in the
depletion layer, let us follow a phenomenological approach suggested
by Sze~\cite{5} and introduce the empirical nonideality factor $n$ so as
to provide the following replacement:
\begin{equation}
\beta\equiv\frac{q}{\kappa T} \to
\frac{q}{n\kappa T}\equiv\beta_n.
\label{eq:53}
\end{equation}

Values of the factor $n$ lie between 1 and 2~\cite{5}: (a)~${n=1}$ when
contribution of the bulk recombination processes is negligibly
small,\, (b)~${n=2}$ when the recombination current dominates over the
diffusion one.

By using the replacement (\ref{eq:53}), expression (\ref{eq:51}) for the
low-frequency dynamic conductance can be rewritten in the following
corrected form:
\begin{equation}
G_d(0,V_\sim)= G_{d0}(V_0)\,g_n(V_{\sim}),
\label{eq:54}
\end{equation}
where the correcting function $g_n(V_{\sim})$ is defined as
\begin{equation}
g_n(V_{\sim}) =
{I_1(\beta_n V_{\sim})\over \beta_n V_{\sim}/2}\equiv
{I_1(\beta V_{\sim}/n)\over \beta V_{\sim}/2n}\,.
\label{eq:55}
\end{equation}

In general, the modified quantity $\beta_n$ in expression (\ref{eq:55})
takes into account not only a contribution from the bulk recombination
processes by means of the nonideality factor $n$ but also that from a priori
unknown temperature $T$ of the $p\!-\!n$-junction under experimental
investigation. Hence, the product $nT$ can be used as a fitting parameter
to adjust the theoretical relations (\ref{eq:54})--(\ref{eq:55}) with
experimental results obtained below.

To verify our theory, we built a simple apparatus using the dual-phase
DSP lock-in amplifier, Stanford Research Systems model SR830. For
the low frequency measurement, the low noise electrometer grade operational
amplifier, Burr-Brown OPA\,128JM, is used.
For the high-frequency measurements, the same lock-in amplifier, combined
with a mixer, and high-frequency op-amps can be used, see Fig.~3.
The AC voltage is  applied to the diode with no bias ($V_0=0$).
The lock-in amplifier has two displays ---
X and Y, which give the root-mean-square (rms) value of output signal
at the excitation frequency~$\omega$. It is easy to show that the lock-in
output in the X and Y displays is equal to
\begin{equation}
V_{out}^X= -RV_{in} \Bigl[
G_d(\omega)\cos(\phi-\phi_0) +
\omega C_d(\omega)\sin(\phi-\phi_0) \Bigr],
\label{eq:56}
\end{equation}
\begin{equation}
V_{out}^Y= -RV_{in} \Bigl[
G_d(\omega)\sin(\phi-\phi_0) +
\omega C_d(\omega)\cos(\phi-\phi_0) \Bigr].
\label{eq:57}
\end{equation}

\begin{figure}
        \epsfysize=2.5in
        \centerline{\epsffile{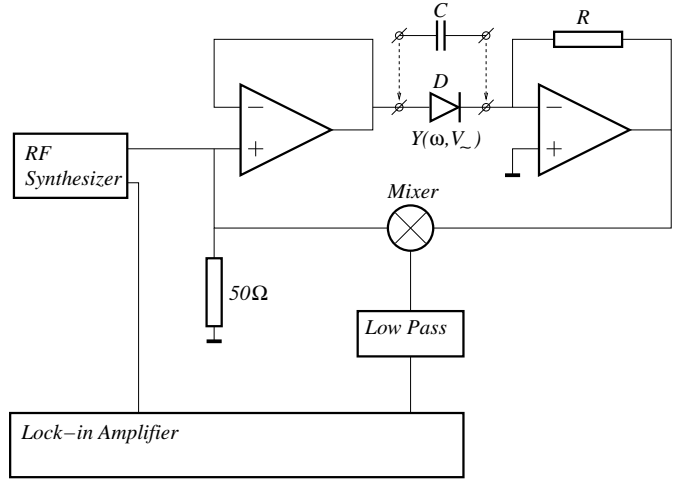}}
\label{Fig3}
\caption{Experimental setup with the measured diode $D$ and reference
capacitor $C$.
}
\end{figure}

Here $R$ is a feedback resistor, $V_{in}$ is the rms value of
the lock-in oscillator voltage ($V_{in}=V_\sim/\sqrt{2}$),
$\phi$ is an internal phase of the lock-in oscillator signal
connected to the phase detector, and $\phi_0$
is a phase of the signal after passing through
an external circuit. From (\ref{eq:56}) and (\ref{eq:57}) it
follows that the two displays are $\pi/2$ out
of phase from each other.

To get correct experimental values, it is necessary to adjust the phase
$\phi$ of the lock-in local oscillator so that the X-display would be used
for the conductance voltage $|V_{out}^X|=RV_{in}G_d$ and the Y-display for
the capacitance voltage $|V_{out}^Y|=RV_{in}\omega C_d$. As follows from
expressions~(\ref{eq:56}) and (\ref{eq:57}), it can be realized only if
$\phi=\phi_0$. The phase $\phi$ is adjusted with the reference capacitor
$C$ which is placed between the local oscillator and the minus input of the
operational amplifier instead of diode~$D$, as shown in Fig.~3. The phase
adjustment is carried out until zero voltage is observed in the
conductance X-display, i.~e., when $\phi=\phi_0$. Then Eqs.~(\ref{eq:56})
and (\ref{eq:57}) yield the required results:
\begin{eqnarray}
G_d(\omega,V_\sim)= {1\over R}\,{|V_{out}^X|\over V_{in}}\,
\biggl|_{\,\phi\,=\,\phi_0},  \qquad
\omega C_d(\omega,V_\sim)= {1\over R}\,{|V_{out}^Y|\over V_{in}}\,
\biggl|_{\,\phi\,=\,\phi_0}.
\label{eq:58}
\end{eqnarray}

After the adjustment, the reference capacitor $C$ is replaced by the
measured diode $D$ and the measurements are performed by varying
the applied AC voltage $V_\sim$ with $V_0=0$. The operating frequency
of the lock-in internal oscillator was chosen 1\,kHz to surely provide
the relation ${\omega\tau_p\ll 1}$ underlying the initial theoretical
expression~(\ref{eq:51}). To extract the conductance, one must
normalize the data.

For our experiments, we have employed a common high frequency diode
1N914B. The experimental results for the above-mentioned diode are presented
in Fig.~4 which demonstrates a quite satisfactory agreement between
the experimental data and our theoretical results, in the low
frequency range.  All resistors and capacitors employed in the experiment
were measured with the Stanford Research Systems model SR720 LCR
meter. The measurement error is found to be below $1\%$. In an additional
paper, the detailed experimental results will be presented.

\begin{figure}
        \epsfysize=3.3in
        \centerline{\epsffile{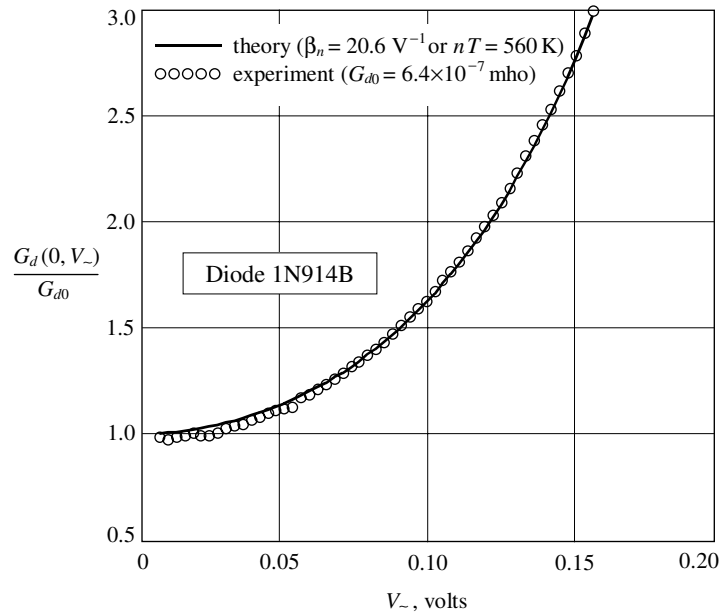}}
\label{Fig4}
\caption{Theoretical and experimental dependencies of the normalized
dynamic conductance $G_d(0,V_\sim)/G_{d0}$ on the peak voltage
$V_\sim$ of AC signal for the 1N914B diode with no DC bias~($V_0=0$).}
\end{figure}

As seen from Eqs.~(\ref{eq:54}) and (\ref{eq:55}), the small voltage
measurement (when $g_n\!=1$) gives a value of the low-frequency dynamic
conductance $G_{d0}$ defined by formula~(\ref{eq:50}). This value has
proved to be equal $G_{d0}\simeq 6.4\times 10^{-7}$\,mho. From here
it follows that $J_s= G_{d0}/\beta_n= 31$ nA,  which is in agreement
with the reverse current for the 1N914B diode given by manufacturers.

From the experimental curve plotted in Fig.~4 it follows that the
modified factor $\beta_n\equiv q/n\kappa T$ should be equal
to 20.6 for the 1N914B diode in order to fit the theoretical
expression~(\ref{eq:54}). Hence, the fitting parameter $nT$ is
equal to $560$\,K, which for the operating temperature ${T=400}$\,K
provides the nonideality factor ${n=1.4}$. Therefore, the measured diode
operates in a regime when the diffusion current slightly dominates
over the recombination current.

\section{Conclusion}

Spectral approach to the theory of $p\!-\!n$-junctions has allowed
us to take into account the effect
of large signal at both the low and high frequencies as compared to
$\tau_{p,n}^{-1}$. This approach is based on the known diffusion
equations for injected minority carriers, which is a standard
practice for semiconductor electronics.

The only specific feature distinguishing our approach from the conventional
one given for comparison in Appendix is related to the initial
representation of the desired carrier concentrations in the
form of Fourier expansion over frequency harmonics. Such harmonics are
produced by nonlinear processes in the $p\!-\!n$-junction when the
sufficiently large AC voltage ${V_1=V_\sim/2}$ is applied to the junction
together with the DC bias voltage~$V_0$. As a result, the spectra of
both the excess concentration of injected carriers and the external
circuit current have been derived (see formulas (\ref{eq:25}),
(\ref{eq:26}), and (\ref{eq:32})). The use of appropriate terms in
Fourier series has given rise to expressions for the DC component~$J_0$
and AC component~$J_1$ of the external circuit current. The former
determine the static current--voltage characteristic $J_0(V_0)$
(formula~(\ref{eq:34})) and from the latter follows the dynamic
admittance ${Y(\omega)=J_1/V_1}$ (formulas~(\ref{eq:37})--(\ref{eq:39})).
These expressions have proved to be dependent not only on $V_0$ but also
on $V_1$, in the case of the nonlinear (in signal) regime of operation
of the $p\!-\!n$-junction.

The results are corroborated by experimental verification, as follows 
from Fig.~4. Detailed experimental results will be presented in a 
separated paper.

\appendix

\section{Conventional Approach to Analysis of the Current--Voltage
Characteristic for $p\!-\!n$-Diodes}

Conventional theory of the $p\!-\!n$-diode operation under the AC signal
of arbitrary amplitude is based on generalizing the known expression for
the static current--voltage characteristic to time-varying conditions:
\begin{equation}
J(t)= J_s \bigl( e^{\,\beta v(t)} - 1 \bigr).
\label{eq:A.1}
\end{equation}
Here the total voltage $v(t)$ applied to the diode includes the DC bias
voltage $V_0$ and the harmonic signal voltage $V_\sim \cos\omega t$
(see Eq.~(\ref{eq:5})). After substituting (\ref{eq:5})
into Eq.~(\ref{eq:A.1}), it~is necessary to use the following Fourier
series~\cite{6} (see also formula 8.511.4 in~\cite{7}):
\begin{eqnarray}
e^{\,\beta V_\sim\cos\omega t} =
I_0(\beta V_\sim) + 2I_1(\beta V_\sim)\cos\omega t +
2I_2(\beta V_\sim)\cos2\omega t+ \ldots,
\label{eq:A.2}
\end{eqnarray}
where $I_0, I_1, I_2,\ldots$ are the modified Bessel functions of
first kind. Insertion of the Fourier series~(\ref{eq:A.2}) into
expression~(\ref{eq:A.1}) yields
\begin{eqnarray}
J(t) &=&
J_s \biggl\{ \Bigl[ I_0(\beta V_\sim) e^{\beta V_0} - \!1 \Bigr] +
2I_1(\beta V_\sim)e^{\beta V_0}\cos\omega t
\nonumber\\[1.5mm]
&& \quad\, +
2I_2(\beta V_\sim) e^{\beta V_0}\cos2\omega t + \ldots\, \biggr\}.
\label{eq:A.3}
\end{eqnarray}

Expression (\ref{eq:A.3}) is the Fourier expansion of the diode current
whose DC component gives the static current--voltage characteristic
\begin{equation}
J_0(V_0,V_{\sim})=
J_s \Bigl[ I_0(\beta V_{\sim}) e^{\beta V_0} -\!1 \Bigr],
\label{eq:A.4}
\end{equation}
which is valid for the harmonic signal of arbitrary amplitude $V_{\sim}$.
The first harmonic of the current (\ref{eq:A.3}) allows one to obtain
the dynamic conductance
\begin{equation}
G_d(V_0,V_{\sim})= \frac{J_s\exp{(\beta V_0)}}{\kappa T/q}\,
\frac{I_1(\beta V_{\sim})}{\beta V_{\sim}/2} \equiv G_{d0}(V_0)\,g_1(V_\sim),
\label{eq:A.5}
\end{equation}
where $G_{d0}(V_0)$ and $g_1(V_\sim)$ are given by formulas
(\ref{eq:46}) and (\ref{eq:43}).

Expression (\ref{eq:A.5}) resulting from the conventional approach
describes solely the low-frequency dynamic conductance $G_d$
(cf. Eq.~(\ref{eq:51})), whose
frequency dependence can be obtained only by using the spectral approach,
as it is done in Sec.~\ref{sec:5}.  Moreover, our spectral approach yields
not only $G_d(\omega)$ but also $C_d(\omega)$ for arbitrary signals.
In the case of small signals, the similar frequency dependencies are
derived in the literature (for example, see~\cite{5}), but they never follow
from the current--voltage characteristic in the form of Eq.~(\ref{eq:A.1})
taken as an initial point for derivation.

\section*{Acknowledgements}

We thank CNPq/``Instituto do Mil\^{e}nio'' Iniative.
One of the authors, AAB, also thanks CNPq for
the support during his stay at UFPE.

\vfill

\end{document}